\documentclass[%
 reprint,
 superscriptaddress,
 amsmath,amssymb,
 aps,
]{revtex4-2}
\usepackage[caption=false]{subfig} 
\usepackage{graphicx}
\usepackage{dcolumn}
\usepackage{bm}
\usepackage{ulem} 

\usepackage{color} 
\usepackage{xcolor}
\begin{document}

\preprint{APS/123-QED}

\title{A magneto-optical trap with millimeter ball lenses}

\author{Cainan S. Nichols}
\affiliation{Department of Physics and Astronomy, Eastern Michigan University, Ypsilanti, Michigan 48197, USA}

\author{Leo M. Nofs}
\affiliation{Department of Physics and Astronomy, Eastern Michigan University, Ypsilanti, Michigan 48197, USA}
\affiliation{Department of Physics, University of Michigan, Ann Arbor, Michigan 48109, USA}

\author{Michael A. Viray}
\affiliation{Department of Physics, University of Michigan, Ann Arbor, Michigan 48109, USA}

\author{Lu Ma}
\affiliation{Department of Physics, University of Michigan, Ann Arbor, Michigan 48109, USA}

\author{Eric Paradis}
\email{eparadis@emich.edu}
\affiliation{Department of Physics and Astronomy, Eastern Michigan University, Ypsilanti, Michigan 48197, USA}

\author{Georg Raithel}
\affiliation{Department of Physics, University of Michigan, Ann Arbor, Michigan 48109, USA}

\date{\today}

\begin{abstract}
We present a magneto-optical trap (MOT) design based on millimeter ball lenses, contained within a metal cube of 0.75$^{\prime \prime}$ side length. We present evidence of trapping approximately $4.2\times 10^5$ of $^{85}$Rb atoms with a number density of $3.2\times 10^9$~$\rm{atoms/cm^3}$ and a loading time of 1.3~s. Measurement and a kinetic laser-cooling model are used to characterize the atom trap design. The design provides several advantages over other types of MOTs: the laser power requirement is low, the small lens and cube sizes allow for miniaturization of MOT applications, and the lack of large-diameter optical beam pathways prevents external blackbody radiation from entering the trapping region.
\end{abstract}

\maketitle

\section{\label{sec:intro} Introduction}

The invention of the magneto-optical trap (MOT) \cite{metcalf, philips98} ushered in a new wave of physics research, as it allowed researchers to cool atoms to previously unattainable temperatures. Since then multiple groups have created new MOT designs that use the same tangible principles but feature different optical configurations, often for the sake of miniaturization, portability, and/or ease of setup. These include grating MOTs \cite{lee13, imhof17}, pyramidal MOTs \cite{lee96, arlt98, vangelyn09, hinton16}, and MOTs with five beams \cite{distefano99}, in addition to variations upon the original six-beam configuration. While the classic six-beam MOT design suffices for most cold-atom experiments, numerous precision measurement and atomic-clock experiments require cold-atom systems that are shielded from external perturbations~\cite{ramos17} such as blackbody radiation. Multiple papers have addressed ways to eliminate blackbody radiation in precision measurements and atomic clocks \cite{golovizin19, xu16, ushijima15}. The six-beam MOT is particularly vulnerable to blackbody radiation because the beams have large cross-sections and the necessary optical apertures subtend a large solid angle with respect to the trap center. This results in uncertainties of the blackbody radiation shift \cite{flambaum16}. Precision measurements and atomic clocks may also be sensitive to external DC electric fields \cite{beloy18}, requiring field-zeroing with electrodes or Faraday shielding. MOT designs with built-in Faraday shielding, including the one described in the present work, provide immediate reduction of the Stark shift. Further applications of such MOTs are found in research that requires well-defined electrode configurations with minimally perturbed symmetry, such as cold-atom-loaded Penning~\cite{gabrielse08, choi08} and Paul traps~\cite{schmid12}.

Here we describe a new MOT design that uses millimeter-sized ball lenses that are held in place in a metal frame of less than 2~cm in diameter, contained inside a vacuum chamber. Ball lenses have well-known optical properties \cite{edmundoptics, kim16} and are typically used for optical tweezers \cite{sasaki97, numata06}, but they are not traditionally used in MOTs. In our design, six independent, narrow, collimated cooling beams ($w_0 \sim 0.6$ mm) pass through the ball lenses, which spread the light into divergent cones reminiscent of light-house beams. A MOT forms at the intersection of the six conical beams. In our paper, we present an experimental realization of a ball lens MOT, along with a computational analysis of trapping performance. The ball-lens MOT design has some specific advantages over the standard six-beam MOT. Because the trap center is enclosed inside of a small metal box, the trapped atoms are Faraday-shielded from stray electric fields. This feature also minimizes perturbations of the blackbody radiation field caused by fields entering through the apertures. Each lens subtends a solid angle of the MOT's surroundings that is on the order of only a few $4 \pi \times 10^{-3}$~steradians. Additionally, the highly divergent nature of the light cones behind the ball lenses makes this MOT operate best at low laser powers. These properties promise great potential for future quantum technology applications \cite{yudin11}. In our work we also describe how specific challenges associated with the beam alignment and geometric peculiarities can be addressed. We further discuss how this style of MOT can be implemented in future experiments.

\section{\label{sec:implementation} Implementation}

\begin{figure}
    \centering
    \subfloat[]{
        \includegraphics[width = 6 cm]{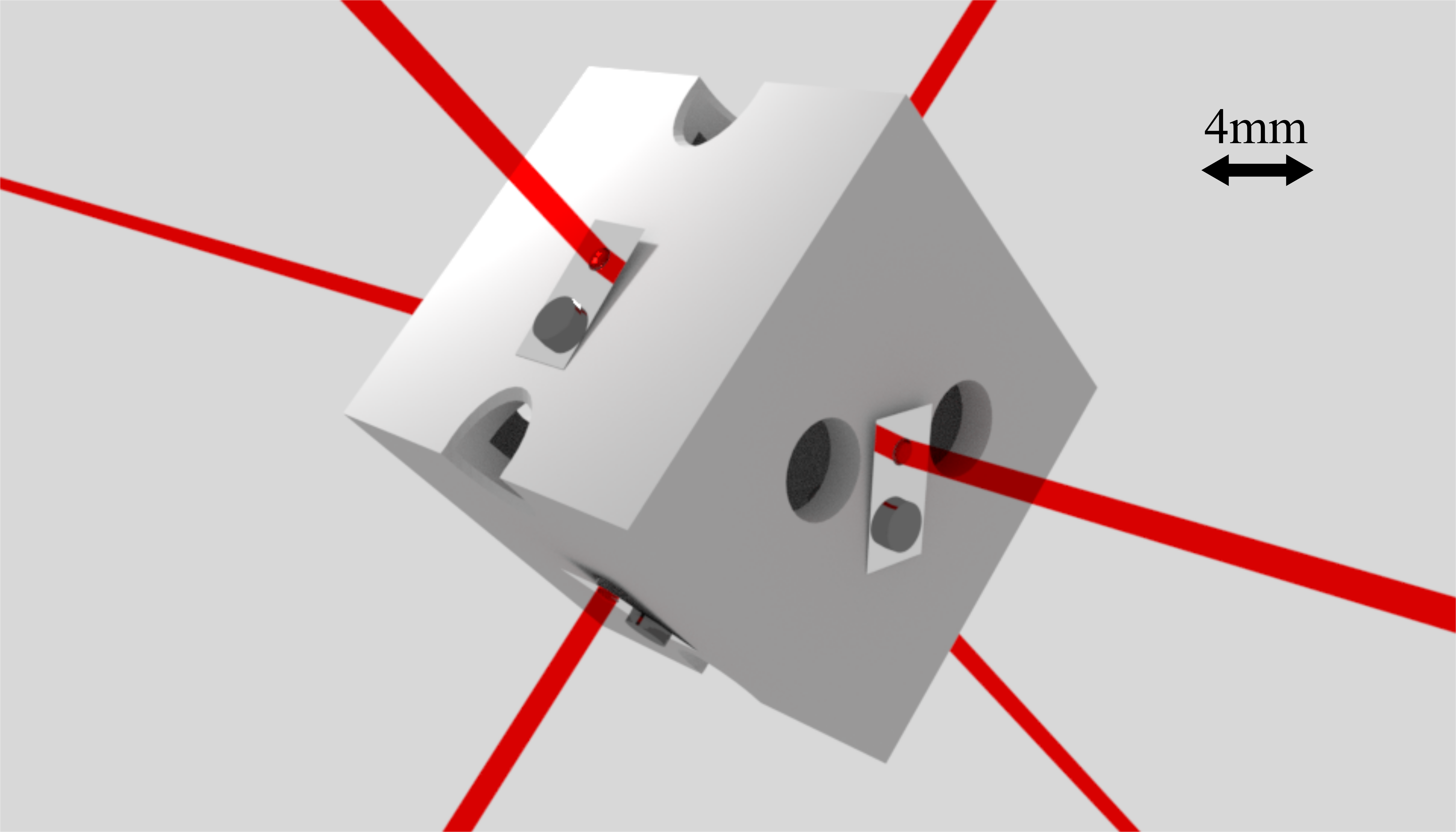}
        \label{subfig:blender}
        }
    \linebreak
     \subfloat[]{
        \includegraphics[width = 6 cm]{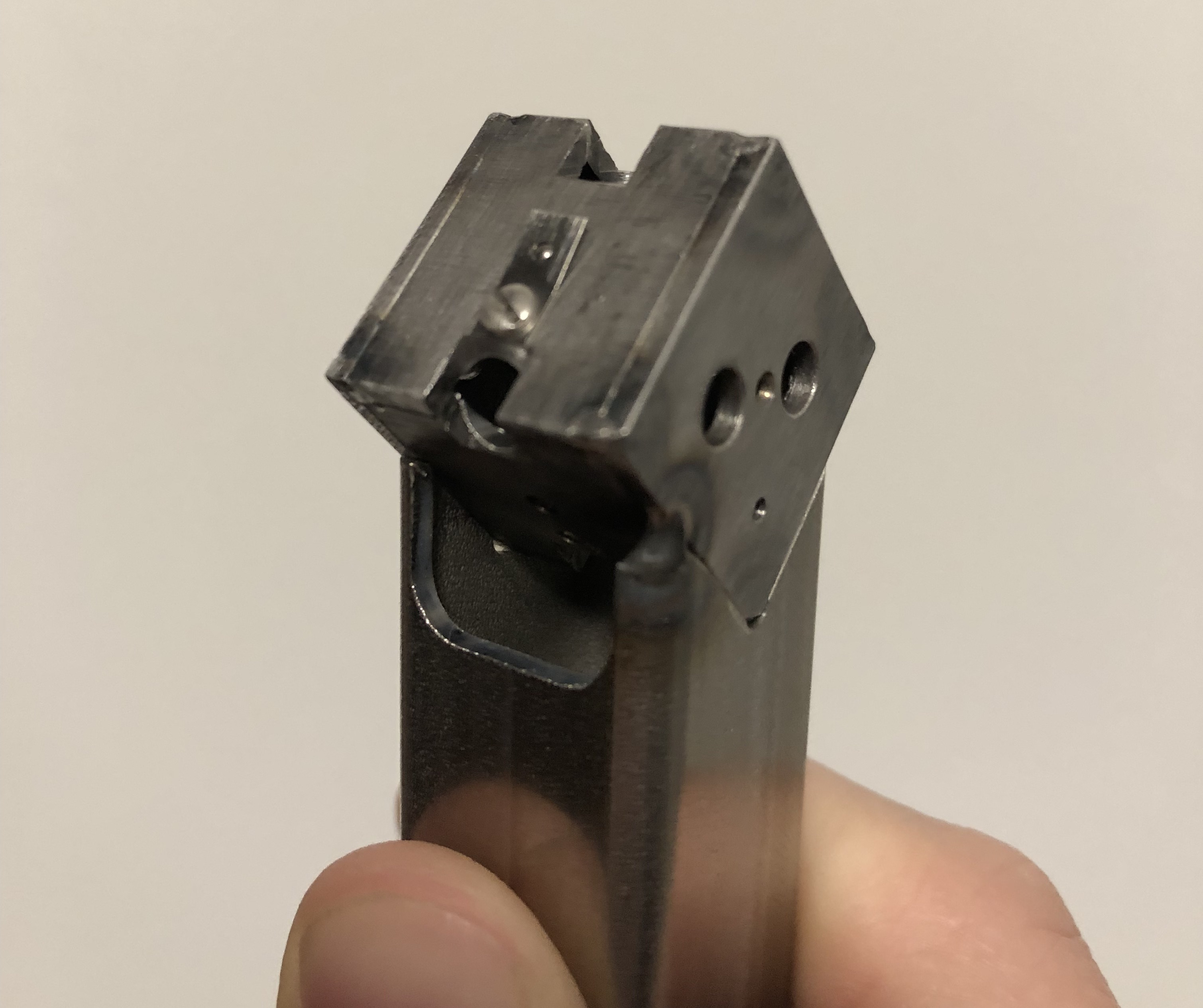}
        \label{subfig:blobpic}
        }
    \linebreak
     \subfloat[]{
        \includegraphics[width = 6 cm]{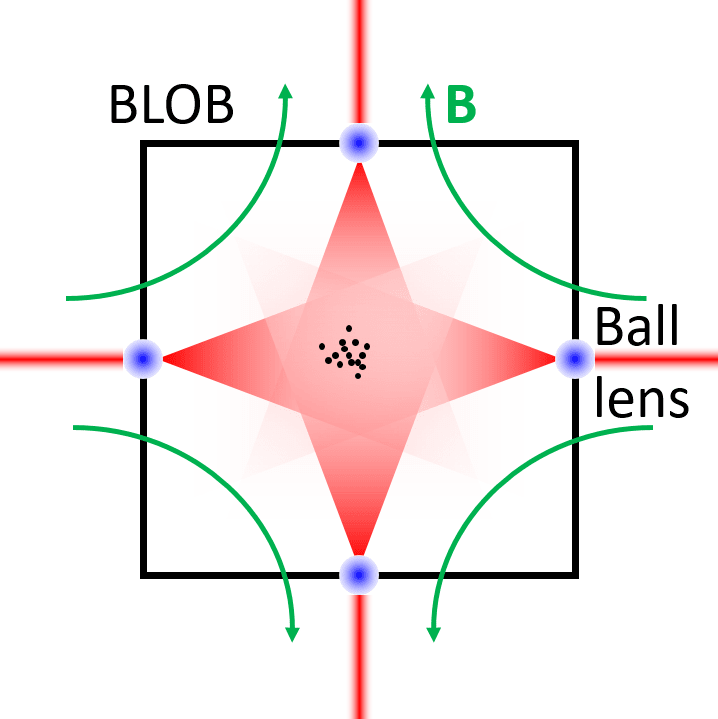}
        \label{subfig:bmot}
        }
    \caption{(a) Computer rendering of ball lens MOT setup, including the cooling beams. The ball lenses are located at the centers of the cube faces. The larger holes are for additional optical access.~(b) Picture of the actual ball lens optical box (BLOB).~(c) Cross-sectional sketch of fields in ball lens MOT. Two of the ball lenses are not included in this drawing.}
    \label{fig:blob}
\end{figure}

In our experimental setup, six 1.5-mm-diameter N-BK7 ball lenses are held in place with a custom-made part called the ball lens optical box (BLOB). Figure~\ref{subfig:blender} shows a computer rendering of the BLOB, and Fig.~\ref{subfig:blobpic} a picture of the physical part. The BLOB is a hollow cube with inner side length of 5/8$^{\prime \prime}$ that is manufactured from six 1/16$^{\prime \prime}$ thick steel plates. Each ball lens is implanted in a counterbore at the center of one of the cube faces, where it is held in place by a metal flap spring. The  flaps have 1.3-mm-diameter holes over the lenses to let light through. The divergent light fields behind the lenses have an approximately Gaussian profile in the directions transverse to the beam axes. The counterbores have inner and outer diameters of 1.3~mm and 2~mm, respectively. The flap springs push the lenses from the outside of the BLOB against 0.5-mm-deep ledges on the inside of the cube faces,
resulting in a simple, secure, and vibration-resistant lens mount. The BLOB cube is welded onto a piece of steel square tubing that is attached to the inside of the vacuum chamber used for MOT testing. The BLOB has eight extra 4-mm-diameter holes on its edges and faces for optical access. While these extra holes are non-essential for MOT operation, they are useful for MOT analysis and other applications involving additional laser beams. 


Unlike in standard 6-beam MOTs, in ball-lens MOTs
it is not practical to re-cycle the laser beams by retro-reflection because of aberrations caused by the lenses. Instead, six independent collimated cooling beams (beam waists $w_0 \sim 0.6$~mm) are directed onto the ball-lenses from the outside of the BLOB. Inside the BLOB, the beams focus into spots at a back focal distance of $\approx 350~\mu$m and emerge as conical beams with a numerical aperture $NA \approx 0.3$, corresponding to an opening angle of the light cones of $\sim 35~^{\circ}$ (full width at half maximum (FWHM) of the intensity). A sketch of the fields inside the BLOB is shown in Fig.~\ref{subfig:bmot}.

The BLOB is tightly mounted between two large-diameter, re-entrant vacuum windows. The two MOT coils are placed close to the outsides of the re-entrant windows, so that the separation between the innermost windings of the coil pair is only $ \sim 1.5^{\prime \prime}$. The MOT coils run at a current of $\sim 6$~A and are air-convection-cooled. The experiment is equipped with bias field coils to adjust the zero position of the MOT quadrupole field.

The ball lens MOT is tested with $^{85}$Rb. The cooling laser is locked to a hyperfine component of the $5S_{1/2} \rightarrow 5P_{3/2}$ transition using saturation spectroscopy in a Rb vapor cell~\cite{demtroder}. The cooling beam passes through an acousto-optic modular (AOM) for switching and for frequency-tuning close to the $F=3 \rightarrow F'=4$ hyperfine cycling transition. A repumper laser drives the $F=2 \rightarrow F'=3$ hyperfine transition. The power of each cooling beam before entering the vacuum chamber is $ \lesssim 5$~mW. Ray tracing analysis of the ball lenses with Zemax shows a waist of the cooling beam at the MOT center of 4~mm, which sets the upper limit of the peak intensity of the individual beams at the MOT center  to $ \sim 12$~$I_{sat}$ (the saturation intensity $I_{sat} =1.6~$mW/cm$^2$ for the cycling transition~\cite{steck2009Rb}). The true intensity is slightly less due to reflection and other losses.

The setup has a probe laser used for shadow imaging. This laser is on-resonance with the $F=3 \rightarrow F'=4$ hyperfine transition and also passes through an AOM for frequency tuning and switching. The MOT probe beam has a FWHM of 2.5~mm and a center intensity to 0.23~mW/cm$^2$. The beam is directed through the vacuum chamber, passed through the center of the BLOB, and aligned into a CCD camera (Pixelfly Model 270 XS). The camera is used to record shadow images, with the probe beam on and MOT beams off, as well as fluorescence images, with the probe beam off and MOT beams on.

\section{\label{analysis} MOT Analysis}

The main quantities of interest in our experimental characterization of the MOT are atom number, atom density, and loading time. The atom number and density are dependent on the design of the MOT and laser parameters, while the loading time is mostly dependent on background pressure and is similar to other MOTs.

\begin{figure}
    \centering
    \subfloat[]{
        \includegraphics[width = 8 cm]{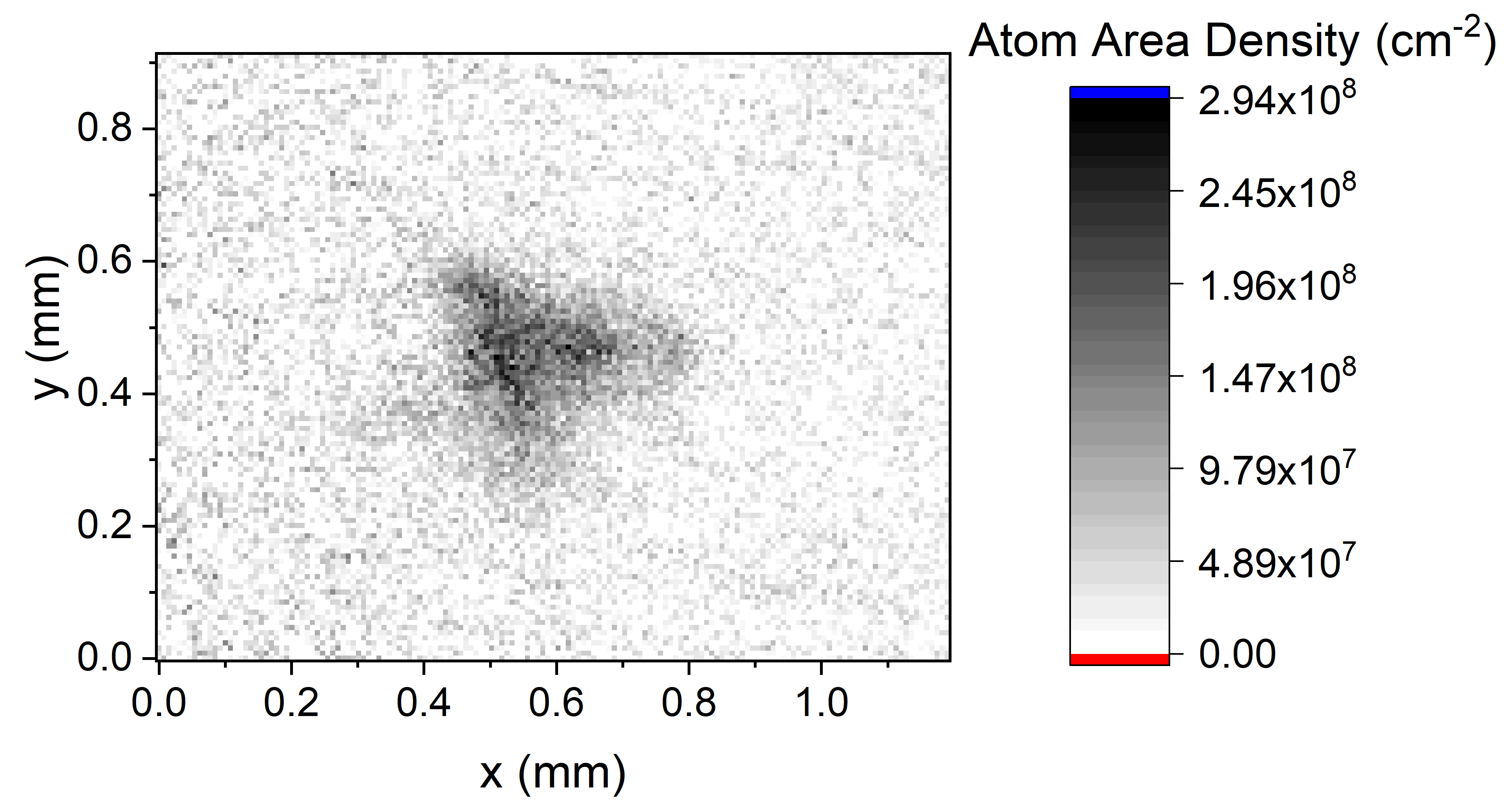}
        \label{subfig:shadow_image}
        }
    \linebreak
     \subfloat[]{
        \includegraphics[width = 6 cm]{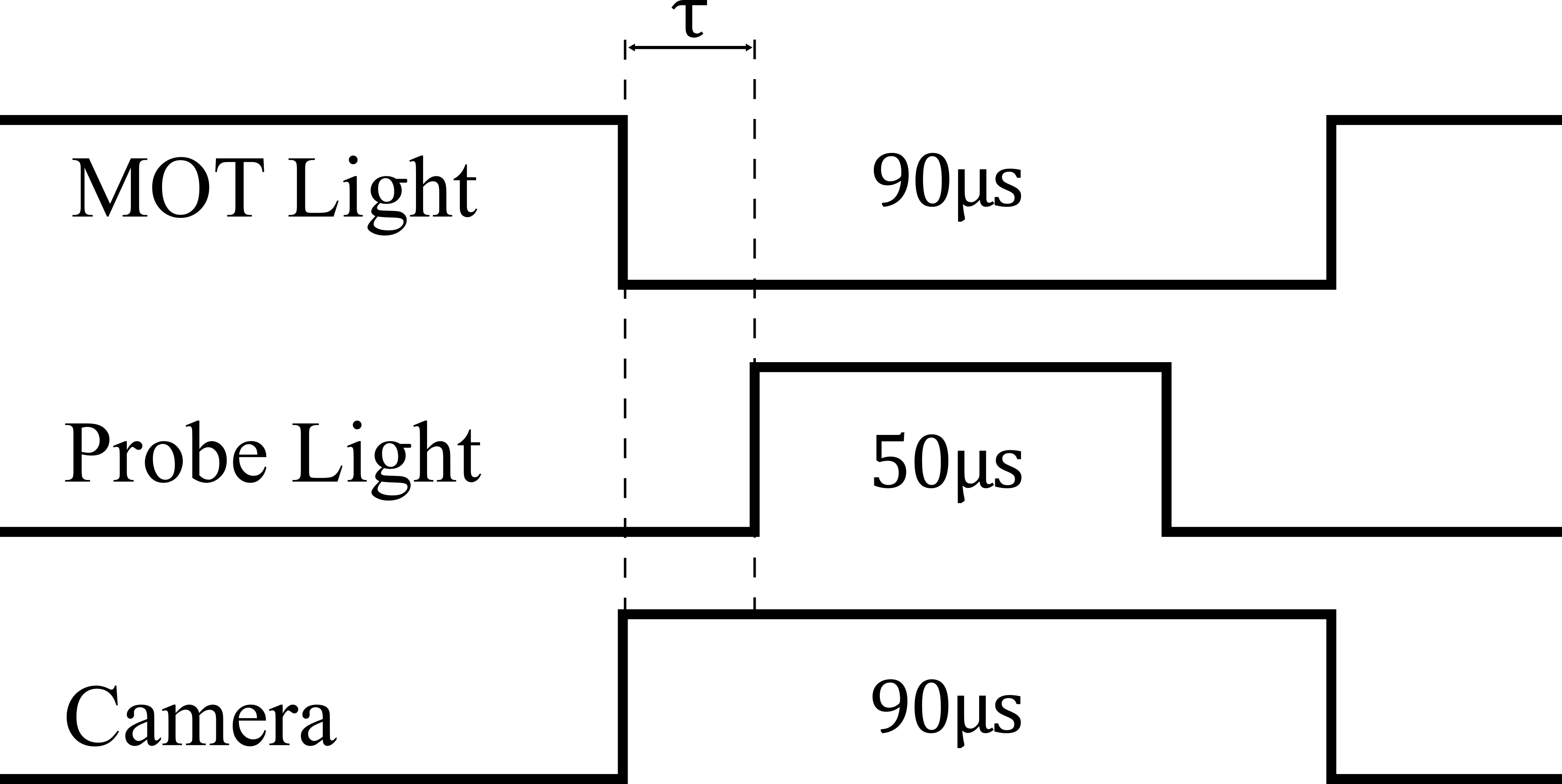}
        \label{subfig:shadow_timing}
        }
    \caption{ (a) A shadow image of the MOT showing area density (the integrand in Eq.~(1)) vs. position in the MOT plane.  (b) Timing diagram for the shadow imaging. When the MOT light is turned off, the camera is gated on for 90~$\mu$s. Following a wait time of $\tau = 20~\mu$s after the MOT light is turned off, the probe pulse is turned on for a duration of 50~$\mu$s.  }
    \label{fig:shadow}
\end{figure}

The atom number, the most important metric in our comparison with theory in Sec.~IV, and the number density of the MOT are measured using shadow imaging. In this configuration, the MOT cooling beams are briefly turned off, and a probe beam pulse is sent through the MOT and into the Pixelfly camera. The MOT and probe beams are switched with the AOMs. Figure~\ref{subfig:shadow_image} shows an area-density shadow image for a MOT single-beam intensity of $I \approx 3.5~$mW/cm$^2$, and Fig.~\ref{subfig:shadow_timing} shows the corresponding timing details. With the probe beam carefully tuned on-resonance with the MOT transition, we find the atom number in our MOT to be 
$N_{MOT, Exp}=4.2\times10^5$.
The value for $N_{MOT, Exp}$ is found by evaluating 

\begin{equation}
N_{MOT, Exp} = \int \frac{1}{\sigma} \ln \left(\frac{I_0(x,y)-I_B(x,y)}{I(x,y)-I_B(x,y)}\right) dx dy
\end{equation}

over the MOT object plane, where $I(x,y)$ is the shadow image with MOT atoms present, $I_B(x,y)$ is
a background image with the probe beam off but all other light sources left on, 
and $I_0(x,y)$ is an image with the probe beam on but without MOT atoms. Since the light is unpolarized and the MOT magnetic field is left on, we use the isotropic absorption cross section $\sigma = 1.246 \times 10^{-9}$~cm$^{2}$~\cite{steck2009Rb}. The integral is evaluated as a discrete sum over a two-dimensional array of CCD pixels, with the pixel area given by the CCD pixel area projected into the object (MOT) plane.  
Assuming that the MOT fills a cubic volume with a side length of 0.5~mm, the typical linear size of the MOT seen from two different observation angles, the atom density in the MOT is estimated to be $3.2 \times10^9$ atoms/cm$^{3}$. 



\begin{figure*}
   \centering
    \includegraphics[width=14 cm]{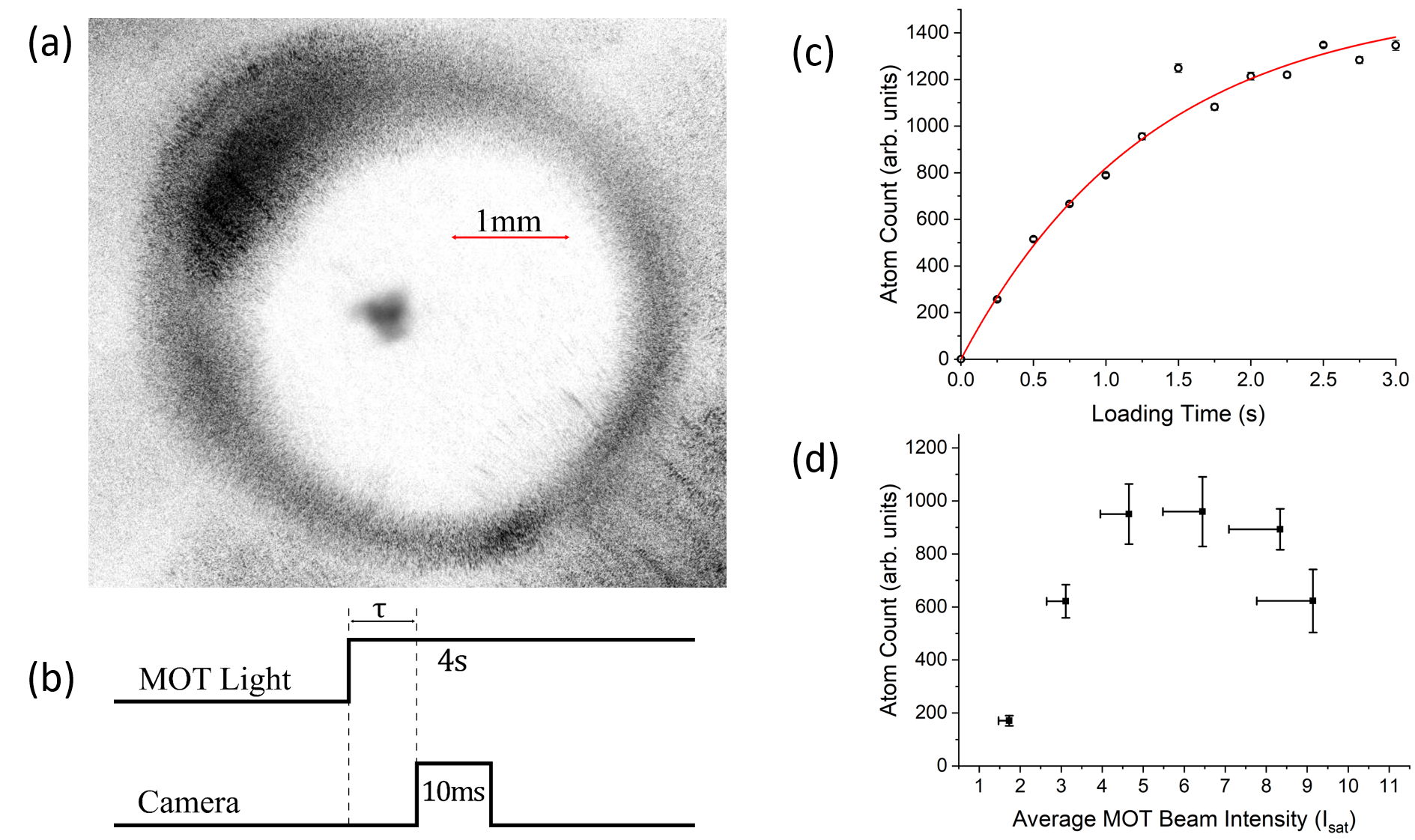}
   \caption{(a) Inverted fluorescence image of the MOT taken through one of the 4-mm viewports in the BLOB with length scale bar. 
   (b) Timing diagram for the loading curve. 
   The loading time  $\tau$ is stepped in units of $0.25$~s, and the camera exposure time is 10~ms. 
  (c) Atom count (relative scale) vs loading time. Applying an exponential fit to the data yields a time constant of $1.3 \pm 0.1$~seconds.  (d) Atom count vs central single-beam intensity at the MOT location. The asymmetric horizontal error bars reflect a potential 15\% reduction of beam intensity due to reflection and other losses, while the vertical error bars are the standard deviations of the data sets for each beam intensity.}
   \label{fig:fluorescence}
\end{figure*}

The loading time, required for the theoretical atom-number estimate in Sec.~\ref{sec:modeling}, is determined with fluorescence imaging. In this configuration, the MOT light is periodically switched on for 4~s to allow the trap to accumulate atoms, and then off for 1~s to empty the trap. The camera records images of the atom cloud at 0.25-s time intervals, ranging from 0~s to 3~s of loading time. Figure~\ref{fig:fluorescence}a shows a fluorescence image of the atom cloud after 3~s of loading, along with a timing diagram in Fig.~\ref{fig:fluorescence}b for the fluorescence imaging procedure. In our analysis, we plot the background-subtracted fluorescence profiles from each of the  images and apply Gaussian fits to these profiles. The atom count is then taken to be
proportional to the areas of the fits at the respective time steps. This method applies to small MOTs, such as ours, which are free of radiation trapping effects, and for which the fluorescence yields an approximately linear measure for atom number. Figure~\ref{fig:fluorescence}c represents relative atom count vs loading time, along with error bars from the fits which represent statistical error. Applying an exponential rise time fit to the data, we calculate a time constant of $1.3 \pm 0.1$~s. The residuals between the data and the fit can be attributed to the effects of intensity and frequency fluctuations of the cooling laser on the experimental data.
The observed loading time is in-line with values typically seen in vapor-cell MOTs.

In our performance evaluation we have also studied the dependence of atom number on MOT beam intensity. In Fig.~\ref{fig:theory}d
we show atom numbers in relative units, obtained from MOT fluorescence, vs central beam intensity. The result indicates best performance at an intensity of $\sim 6 I_{sat}$; at higher intensities the atom number drops quickly. This behavior contrasts with standard six-beam MOTs, in which the atom number keeps increasing to considerably higher intensities before leveling off. Our model, presented next, 
reproduces the observed peculiar intensity dependence of the ball lens MOT. 


\section{\label{sec:modeling} Theoretical Model of Ball Lens MOT}

The objective of our model is to determine the dependence of the number of captured atoms in steady state on ball-lens MOT parameters. Further, we study how the atom capture behavior differs from that of a standard six-beam MOT, and how that translates into differences in the steady-state trapped-atom number and into guidelines for best operating conditions for ball-lens MOTs.

We first outline our kinetic laser-cooling model.
We assume a spherical MOT cell volume of 1.0~cm radius, which has a volume similar to that of the BLOB used in the experiment. 
Thermal atoms are generated at a fixed rate $F_{Sim}$ on the cell wall, at random positions on the cell surface with an  inward velocity distribution known from gas kinetics. The atomic trajectories are propagated with a 4-th order Runge-Kutta routine in which the atoms are subjected to the net radiation pressure force from the six MOT beams. The actual position ${\bf{r}}(t)$ of an atom determines position-dependent beam intensities, beam ${\bf{k}}$-vectors and the local MOT magnetic field. Due to the conical nature of the light fields, the  $\bf{k}$-vectors of a given beam depend on position within the beam and cover an angular range that depends on the numerical aperture (NA).  Along the optical axis of a beam, the transverse FWHM increases linearly with distance from the ball lens, and the intensity drops off as 1/distance$^2$.  Further, the polarization of each beam is locally decomposed into three polarization components (linear, left-handed circular and right-handed circular) relative to the local direction of the magnetic field. Hence, the six MOT beams together give rise to 18 radiation-pressure force components acting on the atom, where each component has the described dependencies on position. Saturation of the assumed $J=0$ to $J=1$ MOT transition (wavelength 780~nm, saturation intensity 1.6~mW/cm$^2$, upper-state decay rate $2 \pi  \times 6$~MHz) is taken into account. To account for spontaneous emission of the atoms on the laser-cooling transition, the program implements photon recoil kicks in random directions, chosen with random numbers. Further, every atom carries a lifetime clock that measures time elapsed after entry of the atom into the simulation. The time elapsed governs the exponential decay of the atom. Physically, an atom decay is a collision of a MOT atom with a fast background gas atom. A collision effectively removes the atom from the simulation. 
Atom decays are implemented using random numbers. The $1/e$ atom decay time is chosen to be 1.3~s, the value observed in our experiment.

\begin{figure*}
    \centering
    \includegraphics[width = 15 cm]{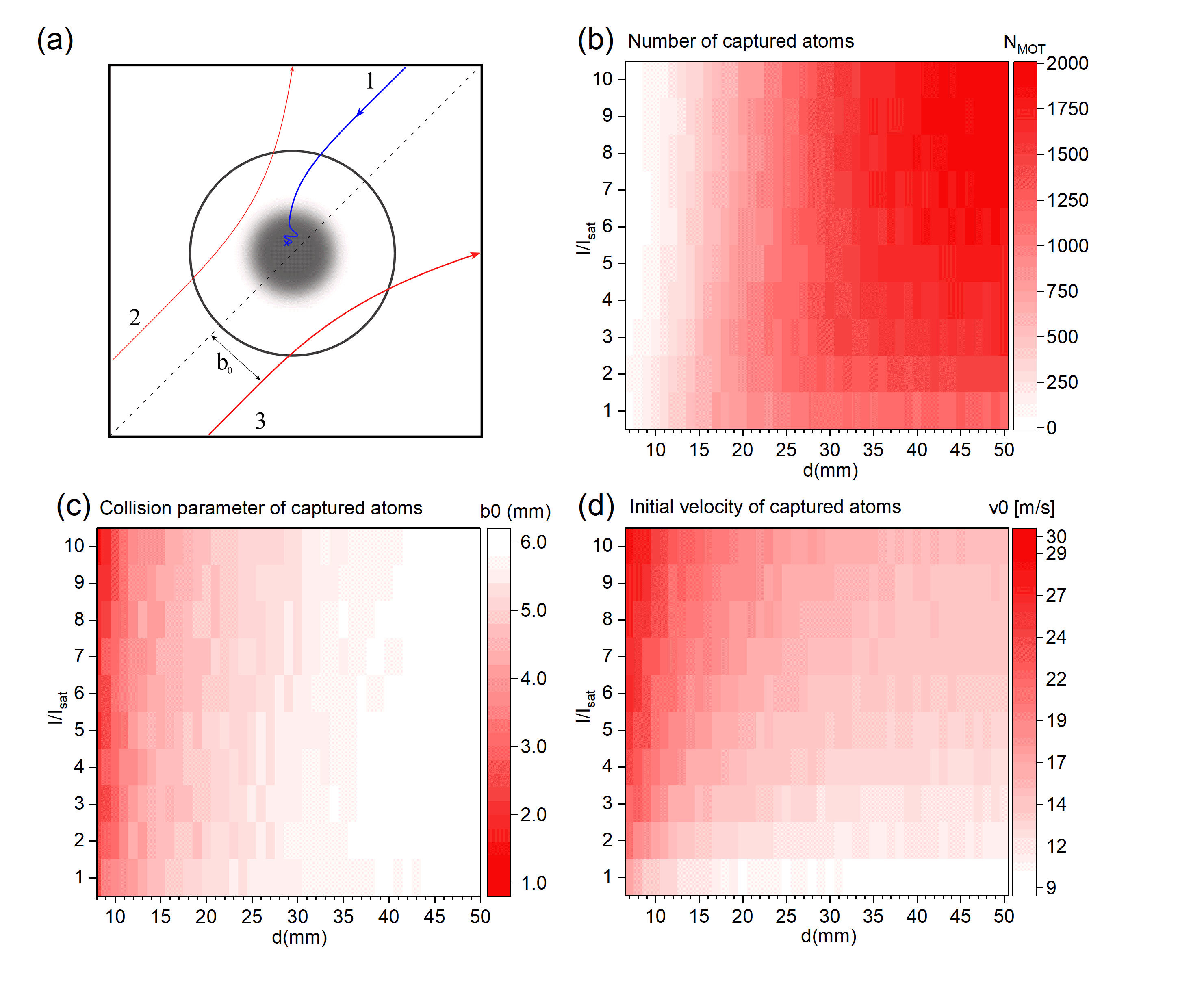}
    \caption{(a) Depiction of three atom trajectories. The relative speeds of the atoms are signified by the thickness of the arrows. Atom~1 is successfully trapped in the MOT. Atom~2 is not trapped because it is moving too slowly, and atom~3 is not trapped because its collision parameter $b_0$ is too large. (b) The number of trapped atoms vs distance $d$ of the ball lenses from the MOT center and $I/I_{sat}$. (c) Root-mean-square (RMS) value of the incident collision parameters $b_0$ of the captured atoms vs $d$ and $I/I_{sat}$. (d) RMS value of the incident speed $v_0$ of the captured atoms vs $d$ and $I/I_{sat}$.}
    \label{fig:theory}
\end{figure*}

Other generic MOT parameters 
for the simulation in Fig.~\ref{fig:theory} include a
MOT agnetic field gradient of 15~G/cm along the field axis and a laser detuning of $-12$~MHz from the MOT transition, corresponding to a typical Rb MOT.
We vary the distance $d$ of the ball lenses from the MOT center from 7~mm to 50~mm.
To study the transition between a regular and the ball lens MOT, we assume that the transverse intensity distributions of the beams are Gaussians with a {\sl{fixed, $d$-independent}} FWHM of 6~mm at the MOT center, as measured for the ball lenses used in the experiment. The FWHM of the beams are  proportional to distance from their respective focal spots. For large $d$, the system approaches a regular 6-beam MOT with near-zero-NA, collimated beams. As $d$ is reduced, the MOT gradually transitions into a ball-lens MOT with large-NA beams. The smaller $d$, the more the divergence of the MOT-beam light cones affects MOT performance. Here we restrict ourselves to the case that all six 
optical axes pass through the center of the MOT, and that all lenses have the same distance from the MOT center.

For each set of parameters we  evaluate the trajectories of $10^7$ to $10^8$ thermal atoms impinging into the MOT region. Every atom that enters gets tagged with its initial speed $v_0$ and collision 
parameter $b_0$ relative to the MOT center (see Fig.~\ref{fig:theory}a). Most atoms do not become trapped (red trajectories), while some do (blue trajectory). Any atom whose speed drops below 1~m/s, within 2~mm from the MOT center, is considered trapped. We have verified that the exact values of these trapping criteria are not important. We log the root-mean-square (RMS) velocity and RMS radius of the trapped-atom cloud in the MOT, the trapped-atom number, the average photon scattering rate of the atoms, and the RMS values of the initial speed $v_0$ and collision parameter $b_0$ of the atoms that become trapped. 


\section{\label{sec:modres} Results of theoretical model}

\subsection{\label{subsec:modres1} Survey of relative performance}


In Fig.~\ref{fig:theory}, we present a survey of ball-lens MOT performance as a function of $d$ and single-beam intensity $I$. The simulated loading time is 0.9~s, the half-life for the collision time constant of 1.3~s measured in the experiment. The impingement flux into the MOT cell, $F_{Sim}$, is taken to be 50 atoms per $\mu$s. (The trapping results are later scaled up to the actual impingement flux.) 
Figure~\ref{fig:theory}a shows a diagram of typical atom trajectories in the MOT, two of which fail and one is successful in becoming trapped. 
Figure~\ref{fig:theory}b displays the number of captured atoms $N_{MOT, Th}$ vs $d$ and ${I}/{I_{Sat}}$. From this simulation it is determined that in the experiment we are operating the MOT near the lower bound in $d$ at which
the ball lens MOT begins to work. At $d \lesssim 7$~mm the ball lens MOT does not capture experimentally useful numbers of atoms. The usual six-beam MOT with collimated beams is near-equivalent with the right margin in Fig.~\ref{fig:theory}b, where $d=50$~mm. The numerical data show that 
at low ${I}/{I_{Sat}}$ our $d \approx 8$~mm ball-lens MOT captures between 10\% of the number of atoms one would find in a regular six-beam MOT, while at high intensity the ratio 
drops to about 1\%. It is also noted that at small $d$-values the performance is best at lower intensities and  degrades  at high intensities. This accords with the trend measured in Fig.~\ref{fig:fluorescence}d. For completeness we also report that the temperature of the trapped atoms is, universally for all cases in Fig.~\ref{fig:theory}, near the Doppler limit (here, $150~\mu$K), and the diameter of the trapped-atom cloud near 0.25~mm. 

In Fig.~\ref{fig:theory}c it is seen that the incident collision parameters $b_0$ of the trapped atoms drops from about 6~mm for the regular MOT ($d=50$~mm) to about 2~mm for the extreme ball-lens MOT with $d=7$~mm, with a minor variation as a function of intensity. Figure~\ref{fig:theory}d shows that for the standard MOT ($d=50$~mm), the RMS capture velocity ranges between 10~m/s at low and 15~m/s at high intensity; this range generally is as expected. Interestingly, for the extreme ball-lens MOT ($d=7$~mm) the RMS capture velocities increase to about  
20~m/s at low and 30~m/s at high intensity. A closer study, not shown, reveals that the $d=7$~mm ball-lens MOT does not capture significant fractions of the slow incident atoms. At first glance this appears counter-intuitive, but we offer an explanation for this in the discussion.

\subsection{\label{subsec:modres2} Quantitative model for the number of trapped atoms}

Next we perform an order-of-magnitude comparison between experimentally observed and simulated trapped-atom numbers. The atom flux impinging into the cell is given by $F_{Exp}=(A/4) n_{V} \bar{v}$, where $A$ is the surface area of the cell that is exposed to impinging thermal atoms, $n_{V}$ is the vapor volume density, and $\bar{v} = \sqrt{8 k_b T/ (\pi M)}$ with cell temperature $T=293$~K and atom mass $M=85$~amu is the  average thermal speed. It is invalid to set $n_{V}$ equal to the room-temperature equilibrium density of Rb, because the MOT cell is not saturated with Rb vapor. For a good comparison it is essential to perform an in-situ reference absorption measurement, from which $n_{V}$ is inferred. 

For the necessary calibration of $n_{V}$, we have performed a reference absorption measurement with a Gaussian-profile test beam diameter $<1~$mm, central intensity 11~mW/cm$^2$, tuned to the
MOT transition) propagating through a 40-cm-long segment of the MOT chamber, with the MOT magnetic field off.  The observed absorption in the vapor cell was $3.6 \pm 0.2\%$. We then apply a numerical absorption model in which we assume a thermal Rb vapor with a Maxwell velocity distribution for 293~K and an unknown volume density $n_{V}$. In the model, the Gaussian beam is segmented into annular rings with given intensities, radii and radial step sizes. The velocity-averaged absorption coefficient, which depends on intensity due to saturation of the transition, is calculated for each ring. The beam power transmitted through the 40-cm long sample is then found by integration over the annular rings. The $n_V$-value
in the calculation is then calibrated to yield the experimentally observed $3.6 \pm 0.2\%$ power loss; the result
is $n_{V}= 6.3 \times 10^{13}~$m$^{-3}$.
With the average thermal speed of $\bar{v}=270$~m/s, the impingement flux density then follows to be
$\bar{v} n_{V}/4 = 4.3 \times 10^{15}~$m$^{-2}$s$^{-1}$. Multiplication of this value with the cross-sectional area $A$ of all apertures leading into the BLOB (eight 4-mm-diameter holes in our experiment) then yields an experimental 
impingement flux $F_{Exp} = A \bar{v} n_{V}/4 = 4.3 \times 10^{11}~$s$^{-1}$.

In the simulation, the impingement flux into the MOT cell is assumed to be $F_{Sim} = 5 \times 10^7~$s$^{-1}$. Therefore, if the simulation shows a trapped-atom number $N_{MOT, Sim}$, the number of trapped atoms expected for our experiment is 
\[N_{MOT, Exp} = N_{MOT, Sim} \frac{F_{Exp}}{F_{Sim}} = 8600 N_{MOT, Sim} \]
We have run the simulation for a best estimate of our experimental conditions ($d=7.5$~mm, beam FWHM of 6~mm, MOT decay time of 1.3~s, strong field gradient 15~G/cm). Good performance is seen in the simulation 
for detuning -15~MHz and $I=2 I_{sat}$, 
where the steady-state atom number is 131. The theoretically predicted value for the trapped atom number then becomes $N_{MOT, Sim}= 1.1 \times 10^6$. This number can be compared with the number of MOT atoms we have experimentally observed under good conditions, $N_{MOT, Exp}= 4.2 \times 10^5$. 

\section{\label{sec:discuss} Discussion}

The picture that emerges from the combined results of the survey study is that extreme ball-lens MOTs, {\sl{i.e.}} MOTs with $d \lesssim 8~$mm and beam NA values $NA \gtrsim 0.3$, only capture atoms that are pointing towards an ``active'' trapping center that is about 4~mm across (gray region in Fig.~\ref{fig:theory}a). Atoms traversing through an outer belt of radiation pressure, ranging in diameter from about 4~mm to the outer reaches of the trapping beams, become blown out on their approach towards the MOT center (trajectory~3 in Fig.~\ref{fig:theory}a). The $d=7$~mm ball-lens MOT therefore only captures atoms with velocities between 20~m/s and 30~m/s and collision parameters less than about 2~mm.
It is particularly noteworthy that atoms slower than about 15~m/s and with collision parameters less than about 2~mm do not become trapped (trajectory~2 in Fig.~\ref{fig:theory}a). 
For an atom to become trapped it has to be fast enough that its inertia carries it through the outer belt of ``bad'' radiation pressure, slow enough that it becomes trapped within the inner region of ``good'' radiation pressure, 
and the trajectory of the incident atom also has to point at the center of the MOT to within about 2~mm tolerance. In Fig.~\ref{fig:theory}a only trajectory~1 meets all criteria.

The agreement between experimental and simulated trapped-atom numbers is within about a factor of three. Considering unknowns such as local versus average thermal atom density, the effect of shadowing of half of the 4-mm holes in the BLOB, beam aberrations, differences between the assumed $J=0 \rightarrow J'=1$ MOT transition and the actual $^{85}$Rb  $J=3 \rightarrow J'=4$ transition, etc., this level of agreement is satisfactory and makes us confident that we have captured the essential physical principles of the ball lens MOT.

\section{\label{conclusion} Conclusion}

We have implemented a magneto-optical trap within a small metallic cube using 1.5-mm diameter ball lenses, and we have developed a kinetic atom-trapping model for the 
ball-lens MOT. Simulated results are in good agreement with our experimental observations and parametrize the range over which this MOT should work well.

While this first design of the BLOB was successful in forming a MOT, there are several changes that we will implement for subsequent versions. The simulations indicate that our ball lens distance $d$ from the trap center was near the lower edge of viability. In view of Fig.~\ref{fig:theory}, for future applications $d$-values in the range of 
2~cm appear very attractive, because they correspond to convenient BLOB box dimensions of about 4~cm side length. Atom numbers in such a ball-lens MOT should be about 50$\%$ of the atom number in a similar-sized regular MOT with collimated beams.

We would be remiss if we did not mention the disadvantages of this MOT design. Unlike standard designs, it is difficult to counter-align the beams that lie along the same axis, making the initial alignment more difficult. Due to the dependence of the central intensity in the MOT region on the incident angle on the ball lens, the setup has a higher sensitivity to relative beam powers than a standard MOT. Nevertheless, once our first ball-lens MOT was observed, 
the design provided a sizable range of stability over both the relative beam powers and angles of incidence for each ball lens.

Now that the viability of a ball lens MOT has been demonstrated, the design may be applied in experiments. The compact BLOB box provides well-defined electrostatic boundary conditions and provides a platform to install additional electrodes. This feature makes the BLOB design attractive for research on cold plasmas generated from trapped-atom clouds, where uncontrolled DC electric fields can be a problem. The BLOB also provides an effective shield against radio-frequency, thermal and optical radiation entering the box, because the solid angle subtended by the ball lenses from the center of the box can be made less than one-thousands of $4 \pi$ and the hole sizes for the ball lenses are only about 1~mm in diameter. This aids in controlling AC shifts, which are a limiting effect in optical-lattice optical clocks and have to be considered in high-precision spectroscopy work with Rydberg atoms~\cite{Farley81,ramos17}.
Further, recent work has been done elsewhere on compact ion traps and trapped-ion laser cooling~\cite{mcmahon20,andelkovic13, goodwin16}. The advantages of the ball-lens MOT would coincide well with some requirements of these types of experiments.

\begin{acknowledgments}
This work was supported by (NSF Grant No. 1707377). We thank David Anderson of Rydberg Technologies, Inc. for valuable discussions.

\end{acknowledgments}

\bibliography{blobbib}

\providecommand{\noopsort}[1]{}\providecommand{\singleletter}[1]{#1}%
\begin{thebibliography}{29}%
\makeatletter
\providecommand \@ifxundefined [1]{%
 \@ifx{#1\undefined}
}%
\providecommand \@ifnum [1]{%
 \ifnum #1\expandafter \@firstoftwo
 \else \expandafter \@secondoftwo
 \fi
}%
\providecommand \@ifx [1]{%
 \ifx #1\expandafter \@firstoftwo
 \else \expandafter \@secondoftwo
 \fi
}%
\providecommand \natexlab [1]{#1}%
\providecommand \enquote  [1]{``#1''}%
\providecommand \bibnamefont  [1]{#1}%
\providecommand \bibfnamefont [1]{#1}%
\providecommand \citenamefont [1]{#1}%
\providecommand \href@noop [0]{\@secondoftwo}%
\providecommand \href [0]{\begingroup \@sanitize@url \@href}%
\providecommand \@href[1]{\@@startlink{#1}\@@href}%
\providecommand \@@href[1]{\endgroup#1\@@endlink}%
\providecommand \@sanitize@url [0]{\catcode `\\12\catcode `\$12\catcode
  `\&12\catcode `\#12\catcode `\^12\catcode `\_12\catcode `\%12\relax}%
\providecommand \@@startlink[1]{}%
\providecommand \@@endlink[0]{}%
\providecommand \url  [0]{\begingroup\@sanitize@url \@url }%
\providecommand \@url [1]{\endgroup\@href {#1}{\urlprefix }}%
\providecommand \urlprefix  [0]{URL }%
\providecommand \Eprint [0]{\href }%
\providecommand \doibase [0]{https://doi.org/}%
\providecommand \selectlanguage [0]{\@gobble}%
\providecommand \bibinfo  [0]{\@secondoftwo}%
\providecommand \bibfield  [0]{\@secondoftwo}%
\providecommand \translation [1]{[#1]}%
\providecommand \BibitemOpen [0]{}%
\providecommand \bibitemStop [0]{}%
\providecommand \bibitemNoStop [0]{.\EOS\space}%
\providecommand \EOS [0]{\spacefactor3000\relax}%
\providecommand \BibitemShut  [1]{\csname bibitem#1\endcsname}%
\let\auto@bib@innerbib\@empty
\bibitem [{\citenamefont {Metcalf}\ and\ \citenamefont {van~der
  Straten}(1999)}]{metcalf}%
  \BibitemOpen
  \bibfield  {author} {\bibinfo {author} {\bibfnamefont {H.~J.}\ \bibnamefont
  {Metcalf}}\ and\ \bibinfo {author} {\bibfnamefont {P.}~\bibnamefont {van~der
  Straten}},\ }\href@noop {} {\emph {\bibinfo {title} {Laser cooling and
  trapping}}}\ (\bibinfo  {publisher} {Springer},\ \bibinfo {year}
  {1999})\BibitemShut {NoStop}%
\bibitem [{\citenamefont {Phillips}(1998)}]{philips98}%
  \BibitemOpen
  \bibfield  {author} {\bibinfo {author} {\bibfnamefont {W.~D.}\ \bibnamefont
  {Phillips}},\ }\bibfield  {title} {\bibinfo {title} {Nobel lecture: Laser
  cooling and trapping of neutral atoms},\ }\href
  {https://doi.org/10.1103/RevModPhys.70.721} {\bibfield  {journal} {\bibinfo
  {journal} {Rev. Mod. Phys.}\ }\textbf {\bibinfo {volume} {70}},\ \bibinfo
  {pages} {721} (\bibinfo {year} {1998})}\BibitemShut {NoStop}%
\bibitem [{\citenamefont {Lee}\ \emph {et~al.}(2013)\citenamefont {Lee},
  \citenamefont {Grover}, \citenamefont {Orozco},\ and\ \citenamefont
  {Rolston}}]{lee13}%
  \BibitemOpen
  \bibfield  {author} {\bibinfo {author} {\bibfnamefont {J.}~\bibnamefont
  {Lee}}, \bibinfo {author} {\bibfnamefont {J.~A.}\ \bibnamefont {Grover}},
  \bibinfo {author} {\bibfnamefont {L.~A.}\ \bibnamefont {Orozco}},\ and\
  \bibinfo {author} {\bibfnamefont {S.~L.}\ \bibnamefont {Rolston}},\
  }\bibfield  {title} {\bibinfo {title} {Sub-doppler cooling of neutral atoms
  in a grating magneto-optical trap},\ }\href
  {https://doi.org/10.1364/JOSAB.30.002869} {\bibfield  {journal} {\bibinfo
  {journal} {J. Opt. Soc. Am. B}\ }\textbf {\bibinfo {volume} {30}},\ \bibinfo
  {pages} {2869} (\bibinfo {year} {2013})}\BibitemShut {NoStop}%
\bibitem [{\citenamefont {Imhof}\ \emph {et~al.}(2017)\citenamefont {Imhof},
  \citenamefont {Stuhl}, \citenamefont {Kasch}, \citenamefont {Kroese},
  \citenamefont {Olson},\ and\ \citenamefont {Squires}}]{imhof17}%
  \BibitemOpen
  \bibfield  {author} {\bibinfo {author} {\bibfnamefont {E.}~\bibnamefont
  {Imhof}}, \bibinfo {author} {\bibfnamefont {B.~K.}\ \bibnamefont {Stuhl}},
  \bibinfo {author} {\bibfnamefont {B.}~\bibnamefont {Kasch}}, \bibinfo
  {author} {\bibfnamefont {B.}~\bibnamefont {Kroese}}, \bibinfo {author}
  {\bibfnamefont {S.~E.}\ \bibnamefont {Olson}},\ and\ \bibinfo {author}
  {\bibfnamefont {M.~B.}\ \bibnamefont {Squires}},\ }\bibfield  {title}
  {\bibinfo {title} {Two-dimensional grating magneto-optical trap},\ }\href
  {https://doi.org/10.1103/PhysRevA.96.033636} {\bibfield  {journal} {\bibinfo
  {journal} {Phys. Rev. A}\ }\textbf {\bibinfo {volume} {96}},\ \bibinfo
  {pages} {033636} (\bibinfo {year} {2017})}\BibitemShut {NoStop}%
\bibitem [{\citenamefont {Lee}\ \emph {et~al.}(1996)\citenamefont {Lee},
  \citenamefont {Kim}, \citenamefont {Noh},\ and\ \citenamefont {Jhe}}]{lee96}%
  \BibitemOpen
  \bibfield  {author} {\bibinfo {author} {\bibfnamefont {K.}~\bibnamefont
  {Lee}}, \bibinfo {author} {\bibfnamefont {J.}~\bibnamefont {Kim}}, \bibinfo
  {author} {\bibfnamefont {H.}~\bibnamefont {Noh}},\ and\ \bibinfo {author}
  {\bibfnamefont {W.}~\bibnamefont {Jhe}},\ }\bibfield  {title} {\bibinfo
  {title} {Single-beam atom trap in a pyramidal and conical hollow mirror},\
  }\href@noop {} {\bibfield  {journal} {\bibinfo  {journal} {Opt. Lett.}\
  }\textbf {\bibinfo {volume} {21}},\ \bibinfo {pages} {1177} (\bibinfo {year}
  {1996})}\BibitemShut {NoStop}%
\bibitem [{\citenamefont {Arlt}\ \emph {et~al.}(1998)\citenamefont {Arlt},
  \citenamefont {Marago}, \citenamefont {Webster}, \citenamefont {Hopkins},\
  and\ \citenamefont {Foot}}]{arlt98}%
  \BibitemOpen
  \bibfield  {author} {\bibinfo {author} {\bibfnamefont {J.}~\bibnamefont
  {Arlt}}, \bibinfo {author} {\bibfnamefont {O.}~\bibnamefont {Marago}},
  \bibinfo {author} {\bibfnamefont {S.}~\bibnamefont {Webster}}, \bibinfo
  {author} {\bibfnamefont {S.}~\bibnamefont {Hopkins}},\ and\ \bibinfo {author}
  {\bibfnamefont {C.}~\bibnamefont {Foot}},\ }\bibfield  {title} {\bibinfo
  {title} {A pyramidal magneto-optical trap as a source of slow atoms},\
  }\href@noop {} {\bibfield  {journal} {\bibinfo  {journal} {Opt. Commun.}\
  }\textbf {\bibinfo {volume} {157}},\ \bibinfo {pages} {303} (\bibinfo {year}
  {1998})}\BibitemShut {NoStop}%
\bibitem [{\citenamefont {Vangeleyn}\ \emph {et~al.}(2009)\citenamefont
  {Vangeleyn}, \citenamefont {Griffin}, \citenamefont {Riis},\ and\
  \citenamefont {Arnold}}]{vangelyn09}%
  \BibitemOpen
  \bibfield  {author} {\bibinfo {author} {\bibfnamefont {M.}~\bibnamefont
  {Vangeleyn}}, \bibinfo {author} {\bibfnamefont {P.~F.}\ \bibnamefont
  {Griffin}}, \bibinfo {author} {\bibfnamefont {E.}~\bibnamefont {Riis}},\ and\
  \bibinfo {author} {\bibfnamefont {A.~S.}\ \bibnamefont {Arnold}},\ }\bibfield
   {title} {\bibinfo {title} {Single-laser, one beam, tetrahedral
  magneto-optical trap},\ }\href {https://doi.org/10.1364/OE.17.013601}
  {\bibfield  {journal} {\bibinfo  {journal} {Opt. Express}\ }\textbf {\bibinfo
  {volume} {17}},\ \bibinfo {pages} {13601} (\bibinfo {year}
  {2009})}\BibitemShut {NoStop}%
\bibitem [{\citenamefont {Hinton}\ \emph {et~al.}(2016)\citenamefont {Hinton},
  \citenamefont {Perea-Ortiz}, \citenamefont {Winch}, \citenamefont {Briggs},
  \citenamefont {Freer}, \citenamefont {Moustoukas}, \citenamefont
  {Powell-Gill}, \citenamefont {Squire}, \citenamefont {Lamb}, \citenamefont
  {Rammeloo}, \citenamefont {Stray}, \citenamefont {Voulazeris}, \citenamefont
  {Zhu}, \citenamefont {Kaushik}, \citenamefont {Lien}, \citenamefont
  {Niggebaum}, \citenamefont {Rodgers}, \citenamefont {Stabrawa}, \citenamefont
  {Boddice}, \citenamefont {Plant}, \citenamefont {Tuckwell}, \citenamefont
  {Bongs}, \citenamefont {Metje},\ and\ \citenamefont {Holynski}}]{hinton16}%
  \BibitemOpen
  \bibfield  {author} {\bibinfo {author} {\bibfnamefont {A.}~\bibnamefont
  {Hinton}}, \bibinfo {author} {\bibfnamefont {M.}~\bibnamefont {Perea-Ortiz}},
  \bibinfo {author} {\bibfnamefont {J.}~\bibnamefont {Winch}}, \bibinfo
  {author} {\bibfnamefont {J.}~\bibnamefont {Briggs}}, \bibinfo {author}
  {\bibfnamefont {S.}~\bibnamefont {Freer}}, \bibinfo {author} {\bibfnamefont
  {D.}~\bibnamefont {Moustoukas}}, \bibinfo {author} {\bibfnamefont
  {S.}~\bibnamefont {Powell-Gill}}, \bibinfo {author} {\bibfnamefont
  {C.}~\bibnamefont {Squire}}, \bibinfo {author} {\bibfnamefont
  {A.}~\bibnamefont {Lamb}}, \bibinfo {author} {\bibfnamefont {C.}~\bibnamefont
  {Rammeloo}}, \bibinfo {author} {\bibfnamefont {B.}~\bibnamefont {Stray}},
  \bibinfo {author} {\bibfnamefont {G.}~\bibnamefont {Voulazeris}}, \bibinfo
  {author} {\bibfnamefont {L.}~\bibnamefont {Zhu}}, \bibinfo {author}
  {\bibfnamefont {A.}~\bibnamefont {Kaushik}}, \bibinfo {author} {\bibfnamefont
  {Y.-H.}\ \bibnamefont {Lien}}, \bibinfo {author} {\bibfnamefont
  {A.}~\bibnamefont {Niggebaum}}, \bibinfo {author} {\bibfnamefont
  {A.}~\bibnamefont {Rodgers}}, \bibinfo {author} {\bibfnamefont
  {A.}~\bibnamefont {Stabrawa}}, \bibinfo {author} {\bibfnamefont
  {D.}~\bibnamefont {Boddice}}, \bibinfo {author} {\bibfnamefont {S.~R.}\
  \bibnamefont {Plant}}, \bibinfo {author} {\bibfnamefont {G.~W.}\ \bibnamefont
  {Tuckwell}}, \bibinfo {author} {\bibfnamefont {K.}~\bibnamefont {Bongs}},
  \bibinfo {author} {\bibfnamefont {N.}~\bibnamefont {Metje}},\ and\ \bibinfo
  {author} {\bibfnamefont {M.}~\bibnamefont {Holynski}},\ }\bibfield  {title}
  {\bibinfo {title} {A portable magneto-optical trap with prospects for atom
  interferometry in civil engineering},\ }\href
  {https://royalsocietypublishing.org/doi/10.1098/rsta.2016.0238} {\bibfield
  {journal} {\bibinfo  {journal} {Philos. Trans. R. Soc. London, Ser. A}\
  }\textbf {\bibinfo {volume} {375}},\ \bibinfo {pages} {20160238} (\bibinfo
  {year} {2016})}\BibitemShut {NoStop}%
\bibitem [{\citenamefont {di~Stefano}\ \emph {et~al.}(1999)\citenamefont
  {di~Stefano}, \citenamefont {Wilkowski}, \citenamefont {M{\"u}ller},\ and\
  \citenamefont {Arimondo}}]{distefano99}%
  \BibitemOpen
  \bibfield  {author} {\bibinfo {author} {\bibfnamefont {A.}~\bibnamefont
  {di~Stefano}}, \bibinfo {author} {\bibfnamefont {D.}~\bibnamefont
  {Wilkowski}}, \bibinfo {author} {\bibfnamefont {J.}~\bibnamefont
  {M{\"u}ller}},\ and\ \bibinfo {author} {\bibfnamefont {E.}~\bibnamefont
  {Arimondo}},\ }\bibfield  {title} {\bibinfo {title} {Five-beam
  magneto-optical trap and optical molasses},\ }\href@noop {} {\bibfield
  {journal} {\bibinfo  {journal} {Applied Physics B}\ }\textbf {\bibinfo
  {volume} {69}},\ \bibinfo {pages} {263} (\bibinfo {year} {1999})}\BibitemShut
  {NoStop}%
\bibitem [{\citenamefont {Ramos}\ \emph {et~al.}(2017)\citenamefont {Ramos},
  \citenamefont {Moore},\ and\ \citenamefont {Raithel}}]{ramos17}%
  \BibitemOpen
  \bibfield  {author} {\bibinfo {author} {\bibfnamefont {A.}~\bibnamefont
  {Ramos}}, \bibinfo {author} {\bibfnamefont {K.}~\bibnamefont {Moore}},\ and\
  \bibinfo {author} {\bibfnamefont {G.}~\bibnamefont {Raithel}},\ }\bibfield
  {title} {\bibinfo {title} {Measuring the rydberg constant using circular
  rydberg atoms in an intensity-modulated optical lattice},\ }\href
  {https://doi.org/10.1103/PhysRevA.96.032513} {\bibfield  {journal} {\bibinfo
  {journal} {Phys. Rev. A}\ }\textbf {\bibinfo {volume} {96}},\ \bibinfo
  {pages} {032513} (\bibinfo {year} {2017})}\BibitemShut {NoStop}%
\bibitem [{\citenamefont {Golovizin}\ \emph {et~al.}(2019)\citenamefont
  {Golovizin}, \citenamefont {Fedorova}, \citenamefont {Tregubov},
  \citenamefont {Sukachev}, \citenamefont {Khabarova}, \citenamefont
  {Sorokin},\ and\ \citenamefont {Kolachevsky}}]{golovizin19}%
  \BibitemOpen
  \bibfield  {author} {\bibinfo {author} {\bibfnamefont {A.}~\bibnamefont
  {Golovizin}}, \bibinfo {author} {\bibfnamefont {E.}~\bibnamefont {Fedorova}},
  \bibinfo {author} {\bibfnamefont {D.}~\bibnamefont {Tregubov}}, \bibinfo
  {author} {\bibfnamefont {D.}~\bibnamefont {Sukachev}}, \bibinfo {author}
  {\bibfnamefont {K.}~\bibnamefont {Khabarova}}, \bibinfo {author}
  {\bibfnamefont {V.}~\bibnamefont {Sorokin}},\ and\ \bibinfo {author}
  {\bibfnamefont {N.}~\bibnamefont {Kolachevsky}},\ }\bibfield  {title}
  {\bibinfo {title} {Inner-shell clock transition in atomic thulium with a
  small blackbody radiation shift},\ }\href
  {https://doi.org/10.1038/s41467-019-09706-9} {\bibfield  {journal} {\bibinfo
  {journal} {Nat. Commun.}\ }\textbf {\bibinfo {volume} {10}},\ \bibinfo
  {pages} {1724} (\bibinfo {year} {2019})}\BibitemShut {NoStop}%
\bibitem [{\citenamefont {Xu}\ and\ \citenamefont {Xu}(2016)}]{xu16}%
  \BibitemOpen
  \bibfield  {author} {\bibinfo {author} {\bibfnamefont {Y.-L.}\ \bibnamefont
  {Xu}}\ and\ \bibinfo {author} {\bibfnamefont {X.-Y.}\ \bibnamefont {Xu}},\
  }\bibfield  {title} {\bibinfo {title} {Analysis of the blackbody-radiation
  shift in an ytterbium optical lattice clock},\ }\href
  {https://doi.org/10.1088/1674-1056/25/10/103202} {\bibfield  {journal}
  {\bibinfo  {journal} {Chinese Physics B}\ }\textbf {\bibinfo {volume} {25}},\
  \bibinfo {pages} {103202} (\bibinfo {year} {2016})}\BibitemShut {NoStop}%
\bibitem [{\citenamefont {Ushijima}\ \emph {et~al.}(2015)\citenamefont
  {Ushijima}, \citenamefont {Takamoto}, \citenamefont {Das}, \citenamefont
  {Ohkubo},\ and\ \citenamefont {Katori}}]{ushijima15}%
  \BibitemOpen
  \bibfield  {author} {\bibinfo {author} {\bibfnamefont {I.}~\bibnamefont
  {Ushijima}}, \bibinfo {author} {\bibfnamefont {M.}~\bibnamefont {Takamoto}},
  \bibinfo {author} {\bibfnamefont {M.}~\bibnamefont {Das}}, \bibinfo {author}
  {\bibfnamefont {T.}~\bibnamefont {Ohkubo}},\ and\ \bibinfo {author}
  {\bibfnamefont {H.}~\bibnamefont {Katori}},\ }\bibfield  {title} {\bibinfo
  {title} {Cryogenic optical lattice clocks},\ }\href
  {https://doi.org/10.1038/nphoton.2015.5} {\bibfield  {journal} {\bibinfo
  {journal} {Nature Photonics}\ }\textbf {\bibinfo {volume} {9}},\ \bibinfo
  {pages} {185} (\bibinfo {year} {2015})}\BibitemShut {NoStop}%
\bibitem [{\citenamefont {Flambaum}\ \emph {et~al.}(2016)\citenamefont
  {Flambaum}, \citenamefont {Porsev},\ and\ \citenamefont
  {Safronova}}]{flambaum16}%
  \BibitemOpen
  \bibfield  {author} {\bibinfo {author} {\bibfnamefont {V.~V.}\ \bibnamefont
  {Flambaum}}, \bibinfo {author} {\bibfnamefont {S.~G.}\ \bibnamefont
  {Porsev}},\ and\ \bibinfo {author} {\bibfnamefont {M.~S.}\ \bibnamefont
  {Safronova}},\ }\bibfield  {title} {\bibinfo {title} {Energy shift due to
  anisotropic blackbody radiation},\ }\href
  {https://doi.org/10.1103/PhysRevA.93.022508} {\bibfield  {journal} {\bibinfo
  {journal} {Phys. Rev. A}\ }\textbf {\bibinfo {volume} {93}},\ \bibinfo
  {pages} {022508} (\bibinfo {year} {2016})}\BibitemShut {NoStop}%
\bibitem [{\citenamefont {Beloy}\ \emph {et~al.}(2018)\citenamefont {Beloy},
  \citenamefont {Zhang}, \citenamefont {McGrew}, \citenamefont {Hinkley},
  \citenamefont {Yoon}, \citenamefont {Nicolodi}, \citenamefont {Fasano},
  \citenamefont {Sch\"affer}, \citenamefont {Brown},\ and\ \citenamefont
  {Ludlow}}]{beloy18}%
  \BibitemOpen
  \bibfield  {author} {\bibinfo {author} {\bibfnamefont {K.}~\bibnamefont
  {Beloy}}, \bibinfo {author} {\bibfnamefont {X.}~\bibnamefont {Zhang}},
  \bibinfo {author} {\bibfnamefont {W.~F.}\ \bibnamefont {McGrew}}, \bibinfo
  {author} {\bibfnamefont {N.}~\bibnamefont {Hinkley}}, \bibinfo {author}
  {\bibfnamefont {T.~H.}\ \bibnamefont {Yoon}}, \bibinfo {author}
  {\bibfnamefont {D.}~\bibnamefont {Nicolodi}}, \bibinfo {author}
  {\bibfnamefont {R.~J.}\ \bibnamefont {Fasano}}, \bibinfo {author}
  {\bibfnamefont {S.~A.}\ \bibnamefont {Sch\"affer}}, \bibinfo {author}
  {\bibfnamefont {R.~C.}\ \bibnamefont {Brown}},\ and\ \bibinfo {author}
  {\bibfnamefont {A.~D.}\ \bibnamefont {Ludlow}},\ }\bibfield  {title}
  {\bibinfo {title} {Faraday-shielded dc stark-shift-free optical lattice
  clock},\ }\href {https://doi.org/10.1103/PhysRevLett.120.183201} {\bibfield
  {journal} {\bibinfo  {journal} {Phys. Rev. Lett.}\ }\textbf {\bibinfo
  {volume} {120}},\ \bibinfo {pages} {183201} (\bibinfo {year}
  {2018})}\BibitemShut {NoStop}%
\bibitem [{\citenamefont {Gabrielse}\ \emph {et~al.}(2008)\citenamefont
  {Gabrielse}, \citenamefont {Larochelle}, \citenamefont {Le~Sage},
  \citenamefont {Levitt}, \citenamefont {Kolthammer}, \citenamefont
  {McConnell}, \citenamefont {Richerme}, \citenamefont {Wrubel}, \citenamefont
  {Speck}, \citenamefont {George}, \citenamefont {Grzonka}, \citenamefont
  {Oelert}, \citenamefont {Sefzick}, \citenamefont {Zhang}, \citenamefont
  {Carew}, \citenamefont {Comeau}, \citenamefont {Hessels}, \citenamefont
  {Storry}, \citenamefont {Weel},\ and\ \citenamefont {Walz}}]{gabrielse08}%
  \BibitemOpen
  \bibfield  {author} {\bibinfo {author} {\bibfnamefont {G.}~\bibnamefont
  {Gabrielse}}, \bibinfo {author} {\bibfnamefont {P.}~\bibnamefont
  {Larochelle}}, \bibinfo {author} {\bibfnamefont {D.}~\bibnamefont {Le~Sage}},
  \bibinfo {author} {\bibfnamefont {B.}~\bibnamefont {Levitt}}, \bibinfo
  {author} {\bibfnamefont {W.~S.}\ \bibnamefont {Kolthammer}}, \bibinfo
  {author} {\bibfnamefont {R.}~\bibnamefont {McConnell}}, \bibinfo {author}
  {\bibfnamefont {P.}~\bibnamefont {Richerme}}, \bibinfo {author}
  {\bibfnamefont {J.}~\bibnamefont {Wrubel}}, \bibinfo {author} {\bibfnamefont
  {A.}~\bibnamefont {Speck}}, \bibinfo {author} {\bibfnamefont {M.~C.}\
  \bibnamefont {George}}, \bibinfo {author} {\bibfnamefont {D.}~\bibnamefont
  {Grzonka}}, \bibinfo {author} {\bibfnamefont {W.}~\bibnamefont {Oelert}},
  \bibinfo {author} {\bibfnamefont {T.}~\bibnamefont {Sefzick}}, \bibinfo
  {author} {\bibfnamefont {Z.}~\bibnamefont {Zhang}}, \bibinfo {author}
  {\bibfnamefont {A.}~\bibnamefont {Carew}}, \bibinfo {author} {\bibfnamefont
  {D.}~\bibnamefont {Comeau}}, \bibinfo {author} {\bibfnamefont {E.~A.}\
  \bibnamefont {Hessels}}, \bibinfo {author} {\bibfnamefont {C.~H.}\
  \bibnamefont {Storry}}, \bibinfo {author} {\bibfnamefont {M.}~\bibnamefont
  {Weel}},\ and\ \bibinfo {author} {\bibfnamefont {J.}~\bibnamefont {Walz}}
  (\bibinfo {collaboration} {ATRAP Collaboration}),\ }\bibfield  {title}
  {\bibinfo {title} {Antihydrogen production within a penning-ioffe trap},\
  }\href {https://doi.org/10.1103/PhysRevLett.100.113001} {\bibfield  {journal}
  {\bibinfo  {journal} {Phys. Rev. Lett.}\ }\textbf {\bibinfo {volume} {100}},\
  \bibinfo {pages} {113001} (\bibinfo {year} {2008})}\BibitemShut {NoStop}%
\bibitem [{\citenamefont {Choi}\ \emph {et~al.}(2008)\citenamefont {Choi},
  \citenamefont {Knuffman}, \citenamefont {Zhang}, \citenamefont {Povilus},\
  and\ \citenamefont {Raithel}}]{choi08}%
  \BibitemOpen
  \bibfield  {author} {\bibinfo {author} {\bibfnamefont {J.-H.}\ \bibnamefont
  {Choi}}, \bibinfo {author} {\bibfnamefont {B.}~\bibnamefont {Knuffman}},
  \bibinfo {author} {\bibfnamefont {X.~H.}\ \bibnamefont {Zhang}}, \bibinfo
  {author} {\bibfnamefont {A.~P.}\ \bibnamefont {Povilus}},\ and\ \bibinfo
  {author} {\bibfnamefont {G.}~\bibnamefont {Raithel}},\ }\bibfield  {title}
  {\bibinfo {title} {Trapping and evolution dynamics of ultracold two-component
  plasmas},\ }\href {https://doi.org/10.1103/PhysRevLett.100.175002} {\bibfield
   {journal} {\bibinfo  {journal} {Phys. Rev. Lett.}\ }\textbf {\bibinfo
  {volume} {100}},\ \bibinfo {pages} {175002} (\bibinfo {year}
  {2008})}\BibitemShut {NoStop}%
\bibitem [{\citenamefont {Schmid}\ \emph {et~al.}(2012)\citenamefont {Schmid},
  \citenamefont {H{\"a}rter}, \citenamefont {Frisch}, \citenamefont {Hoinka},\
  and\ \citenamefont {Denschlag}}]{schmid12}%
  \BibitemOpen
  \bibfield  {author} {\bibinfo {author} {\bibfnamefont {S.}~\bibnamefont
  {Schmid}}, \bibinfo {author} {\bibfnamefont {A.}~\bibnamefont {H{\"a}rter}},
  \bibinfo {author} {\bibfnamefont {A.}~\bibnamefont {Frisch}}, \bibinfo
  {author} {\bibfnamefont {S.}~\bibnamefont {Hoinka}},\ and\ \bibinfo {author}
  {\bibfnamefont {J.~H.}\ \bibnamefont {Denschlag}},\ }\bibfield  {title}
  {\bibinfo {title} {An apparatus for immersing trapped ions into an ultracold
  gas of neutral atoms},\ }\href {https://doi.org/10.1063/1.4718356} {\bibfield
   {journal} {\bibinfo  {journal} {Review of Scientific Instruments}\ }\textbf
  {\bibinfo {volume} {83}},\ \bibinfo {pages} {053108} (\bibinfo {year}
  {2012})}\BibitemShut {NoStop}%
\bibitem [{\citenamefont {EdmundOptics}()}]{edmundoptics}%
  \BibitemOpen
  \bibfield  {author} {\bibinfo {author} {\bibnamefont {EdmundOptics}},\ }\href
  {https://www.edmundoptics.com/knowledge-center/application-notes/optics/understanding-ball-lenses/}
  {\bibinfo {title} {Understanding ball lenses}}\BibitemShut {NoStop}%
\bibitem [{\citenamefont {Kim}\ \emph {et~al.}(2016)\citenamefont {Kim},
  \citenamefont {Scharf}, \citenamefont {M\"{u}hlig}, \citenamefont {Fruhnert},
  \citenamefont {Rockstuhl}, \citenamefont {Bitterli}, \citenamefont {Noell},
  \citenamefont {Voelkel},\ and\ \citenamefont {Herzig}}]{kim16}%
  \BibitemOpen
  \bibfield  {author} {\bibinfo {author} {\bibfnamefont {M.-S.}\ \bibnamefont
  {Kim}}, \bibinfo {author} {\bibfnamefont {T.}~\bibnamefont {Scharf}},
  \bibinfo {author} {\bibfnamefont {S.}~\bibnamefont {M\"{u}hlig}}, \bibinfo
  {author} {\bibfnamefont {M.}~\bibnamefont {Fruhnert}}, \bibinfo {author}
  {\bibfnamefont {C.}~\bibnamefont {Rockstuhl}}, \bibinfo {author}
  {\bibfnamefont {R.}~\bibnamefont {Bitterli}}, \bibinfo {author}
  {\bibfnamefont {W.}~\bibnamefont {Noell}}, \bibinfo {author} {\bibfnamefont
  {R.}~\bibnamefont {Voelkel}},\ and\ \bibinfo {author} {\bibfnamefont {H.~P.}\
  \bibnamefont {Herzig}},\ }\bibfield  {title} {\bibinfo {title} {Refraction
  limit of miniaturized optical systems: a ball-lens example},\ }\href
  {https://doi.org/10.1364/OE.24.006996} {\bibfield  {journal} {\bibinfo
  {journal} {Opt. Express}\ }\textbf {\bibinfo {volume} {24}},\ \bibinfo
  {pages} {6996} (\bibinfo {year} {2016})}\BibitemShut {NoStop}%
\bibitem [{\citenamefont {Sasaki}\ \emph {et~al.}(1997)\citenamefont {Sasaki},
  \citenamefont {Kurosawa},\ and\ \citenamefont {Hane}}]{sasaki97}%
  \BibitemOpen
  \bibfield  {author} {\bibinfo {author} {\bibfnamefont {M.}~\bibnamefont
  {Sasaki}}, \bibinfo {author} {\bibfnamefont {T.}~\bibnamefont {Kurosawa}},\
  and\ \bibinfo {author} {\bibfnamefont {K.}~\bibnamefont {Hane}},\ }\bibfield
  {title} {\bibinfo {title} {Micro-objective manipulated with optical
  tweezers},\ }\href {https://doi.org/10.1063/1.118260} {\bibfield  {journal}
  {\bibinfo  {journal} {Appl. Phys. Lett.}\ }\textbf {\bibinfo {volume} {70}},\
  \bibinfo {pages} {785} (\bibinfo {year} {1997})},\ \Eprint
  {https://arxiv.org/abs/https://doi.org/10.1063/1.118260}
  {https://doi.org/10.1063/1.118260} \BibitemShut {NoStop}%
\bibitem [{\citenamefont {Numata}\ \emph {et~al.}(2006)\citenamefont {Numata},
  \citenamefont {Takayanagi}, \citenamefont {Otani},\ and\ \citenamefont
  {Umeda}}]{numata06}%
  \BibitemOpen
  \bibfield  {author} {\bibinfo {author} {\bibfnamefont {T.}~\bibnamefont
  {Numata}}, \bibinfo {author} {\bibfnamefont {A.}~\bibnamefont {Takayanagi}},
  \bibinfo {author} {\bibfnamefont {Y.}~\bibnamefont {Otani}},\ and\ \bibinfo
  {author} {\bibfnamefont {N.}~\bibnamefont {Umeda}},\ }\bibfield  {title}
  {\bibinfo {title} {Manipulation of metal nanoparticles using fiber-optic
  laser tweezers with a microspherical focusing lens},\ }\href
  {https://doi.org/10.1143/jjap.45.359} {\bibfield  {journal} {\bibinfo
  {journal} {Japanese Journal of Applied Physics}\ }\textbf {\bibinfo {volume}
  {45}},\ \bibinfo {pages} {359} (\bibinfo {year} {2006})}\BibitemShut
  {NoStop}%
\bibitem [{\citenamefont {Yudin}\ \emph {et~al.}(2011)\citenamefont {Yudin},
  \citenamefont {Taichenachev}, \citenamefont {Okhapkin}, \citenamefont
  {Bagayev}, \citenamefont {Tamm}, \citenamefont {Peik}, \citenamefont
  {Huntemann}, \citenamefont {Mehlst\"aubler},\ and\ \citenamefont
  {Riehle}}]{yudin11}%
  \BibitemOpen
  \bibfield  {author} {\bibinfo {author} {\bibfnamefont {V.~I.}\ \bibnamefont
  {Yudin}}, \bibinfo {author} {\bibfnamefont {A.~V.}\ \bibnamefont
  {Taichenachev}}, \bibinfo {author} {\bibfnamefont {M.~V.}\ \bibnamefont
  {Okhapkin}}, \bibinfo {author} {\bibfnamefont {S.~N.}\ \bibnamefont
  {Bagayev}}, \bibinfo {author} {\bibfnamefont {C.}~\bibnamefont {Tamm}},
  \bibinfo {author} {\bibfnamefont {E.}~\bibnamefont {Peik}}, \bibinfo {author}
  {\bibfnamefont {N.}~\bibnamefont {Huntemann}}, \bibinfo {author}
  {\bibfnamefont {T.~E.}\ \bibnamefont {Mehlst\"aubler}},\ and\ \bibinfo
  {author} {\bibfnamefont {F.}~\bibnamefont {Riehle}},\ }\bibfield  {title}
  {\bibinfo {title} {Atomic clocks with suppressed blackbody radiation shift},\
  }\href {https://doi.org/10.1103/PhysRevLett.107.030801} {\bibfield  {journal}
  {\bibinfo  {journal} {Phys. Rev. Lett.}\ }\textbf {\bibinfo {volume} {107}},\
  \bibinfo {pages} {030801} (\bibinfo {year} {2011})}\BibitemShut {NoStop}%
\bibitem [{\citenamefont {Demtr{\"o}der}(2008)}]{demtroder}%
  \BibitemOpen
  \bibfield  {author} {\bibinfo {author} {\bibfnamefont {W.}~\bibnamefont
  {Demtr{\"o}der}},\ }\href@noop {} {\emph {\bibinfo {title} {Laser
  Spectroscopy}}}\ (\bibinfo  {publisher} {Springer},\ \bibinfo {year}
  {2008})\BibitemShut {NoStop}%
\bibitem [{\citenamefont {Steck}()}]{steck2009Rb}%
  \BibitemOpen
  \bibfield  {author} {\bibinfo {author} {\bibfnamefont {D.~A.}\ \bibnamefont
  {Steck}},\ }\href@noop {} {\bibinfo {title} {{Rubidium 85 D line data}}},\
  \bibinfo {howpublished} {Available online at {\url{http://steck.
  us/alkalidata}} (Revision 2.1.6, 20 September 2013)}\BibitemShut {NoStop}%
\bibitem [{\citenamefont {Farley}\ and\ \citenamefont {Wing}(1981)}]{Farley81}%
  \BibitemOpen
  \bibfield  {author} {\bibinfo {author} {\bibfnamefont {J.~W.}\ \bibnamefont
  {Farley}}\ and\ \bibinfo {author} {\bibfnamefont {W.~H.}\ \bibnamefont
  {Wing}},\ }\bibfield  {title} {\bibinfo {title} {Accurate calculation of
  dynamic stark shifts and depopulation rates of rydberg energy levels induced
  by blackbody radiation. hydrogen, helium, and alkali-metal atoms},\ }\href
  {https://doi.org/10.1103/PhysRevA.23.2397} {\bibfield  {journal} {\bibinfo
  {journal} {Phys. Rev. A}\ }\textbf {\bibinfo {volume} {23}},\ \bibinfo
  {pages} {2397} (\bibinfo {year} {1981})}\BibitemShut {NoStop}%
\bibitem [{\citenamefont {McMahon}\ \emph {et~al.}(2020)\citenamefont
  {McMahon}, \citenamefont {Volin}, \citenamefont {Rellergert},\ and\
  \citenamefont {Sawyer}}]{mcmahon20}%
  \BibitemOpen
  \bibfield  {author} {\bibinfo {author} {\bibfnamefont {B.~J.}\ \bibnamefont
  {McMahon}}, \bibinfo {author} {\bibfnamefont {C.}~\bibnamefont {Volin}},
  \bibinfo {author} {\bibfnamefont {W.~G.}\ \bibnamefont {Rellergert}},\ and\
  \bibinfo {author} {\bibfnamefont {B.~C.}\ \bibnamefont {Sawyer}},\ }\bibfield
   {title} {\bibinfo {title} {Doppler-cooled ions in a compact reconfigurable
  penning trap},\ }\href {https://doi.org/10.1103/PhysRevA.101.013408}
  {\bibfield  {journal} {\bibinfo  {journal} {Phys. Rev. A}\ }\textbf {\bibinfo
  {volume} {101}},\ \bibinfo {pages} {013408} (\bibinfo {year}
  {2020})}\BibitemShut {NoStop}%
\bibitem [{\citenamefont {Andelkovic}\ \emph {et~al.}(2013)\citenamefont
  {Andelkovic}, \citenamefont {Cazan}, \citenamefont {N\"ortersh\"auser},
  \citenamefont {Bharadia}, \citenamefont {Segal}, \citenamefont {Thompson},
  \citenamefont {J\"ohren}, \citenamefont {Vollbrecht}, \citenamefont
  {Hannen},\ and\ \citenamefont {Vogel}}]{andelkovic13}%
  \BibitemOpen
  \bibfield  {author} {\bibinfo {author} {\bibfnamefont {Z.}~\bibnamefont
  {Andelkovic}}, \bibinfo {author} {\bibfnamefont {R.}~\bibnamefont {Cazan}},
  \bibinfo {author} {\bibfnamefont {W.}~\bibnamefont {N\"ortersh\"auser}},
  \bibinfo {author} {\bibfnamefont {S.}~\bibnamefont {Bharadia}}, \bibinfo
  {author} {\bibfnamefont {D.~M.}\ \bibnamefont {Segal}}, \bibinfo {author}
  {\bibfnamefont {R.~C.}\ \bibnamefont {Thompson}}, \bibinfo {author}
  {\bibfnamefont {R.}~\bibnamefont {J\"ohren}}, \bibinfo {author}
  {\bibfnamefont {J.}~\bibnamefont {Vollbrecht}}, \bibinfo {author}
  {\bibfnamefont {V.}~\bibnamefont {Hannen}},\ and\ \bibinfo {author}
  {\bibfnamefont {M.}~\bibnamefont {Vogel}},\ }\bibfield  {title} {\bibinfo
  {title} {Laser cooling of externally produced mg ions in a penning trap for
  sympathetic cooling of highly charged ions},\ }\href
  {https://doi.org/10.1103/PhysRevA.87.033423} {\bibfield  {journal} {\bibinfo
  {journal} {Phys. Rev. A}\ }\textbf {\bibinfo {volume} {87}},\ \bibinfo
  {pages} {033423} (\bibinfo {year} {2013})}\BibitemShut {NoStop}%
\bibitem [{\citenamefont {Goodwin}\ \emph {et~al.}(2016)\citenamefont
  {Goodwin}, \citenamefont {Stutter}, \citenamefont {Thompson},\ and\
  \citenamefont {Segal}}]{goodwin16}%
  \BibitemOpen
  \bibfield  {author} {\bibinfo {author} {\bibfnamefont {J.~F.}\ \bibnamefont
  {Goodwin}}, \bibinfo {author} {\bibfnamefont {G.}~\bibnamefont {Stutter}},
  \bibinfo {author} {\bibfnamefont {R.~C.}\ \bibnamefont {Thompson}},\ and\
  \bibinfo {author} {\bibfnamefont {D.~M.}\ \bibnamefont {Segal}},\ }\bibfield
  {title} {\bibinfo {title} {Resolved-sideband laser cooling in a penning
  trap},\ }\href {https://doi.org/10.1103/PhysRevLett.116.143002} {\bibfield
  {journal} {\bibinfo  {journal} {Phys. Rev. Lett.}\ }\textbf {\bibinfo
  {volume} {116}},\ \bibinfo {pages} {143002} (\bibinfo {year}
  {2016})}\BibitemShut {NoStop}%
\end{thebibliography}%

\end{document}